\documentclass[aps,onecolumn,twoside,secnumarabic,12pt,balancelastpage,amsmath,amssymb,nofootinbib,hyperref=pdftex]{revtex4}

\usepackage{color}         % produces boxes or entire pages with colored backgrounds
\usepackage{multirow}
\usepackage{graphics}      % standard graphics specifications
\usepackage[pdftex]{graphicx}      % alternative graphics specifications
\usepackage{longtable}     % helps with long table options
\usepackage[english]{babel}
\usepackage{amsmath}
\usepackage{epsf}          % old package handles encapsulated post script issues
\usepackage{bm}            % special 'bold-math' package
\usepackage{verbatim}			% for comment environment
\usepackage[colorlinks=true]{hyperref}  % this package should be added after all others % use as follows: \url{http://web.mit.edu/8.13}

\begin{document}

\begin{flushright}
\today
\end{flushright}

\vspace{0.07in}

\noindent
\begin{center}

{\bf\large Axion-Like Particles at the ILC Giga-Z}

\vspace{0.5cm}
{Noah Steinberg, James D. Wells}

{\it Leinweber Center for Theoretical Physics \\
Physics Department, University of Michigan \\
Ann Arbor, MI 48109-1040 USA}\\
\end{center}

\noindent
{\it Abstract:}
Axion-Like Particles (ALPs) are a generic, calculable, and well motivated extension of the Standard Model with far reaching phenomenology. ALPs that couple only to hypercharge represent one subset of such models, coupling the ALP to both photons and the $Z$ boson. We examine the current constraints on this class of models with an ALP mass in the 100 MeV to 100 GeV range, paying particular attention to the region between 100 MeV to 10 GeV, a portion of parameter space which is ill constrained by current experiments. We show that the more than $10^{9}$ $Z$ bosons produced in the Giga-Z mode of the future ILC experiment, combined with the highly granular nature of its detectors, will allow for ALPs coupled to hypercharge to be discovered with couplings down to nearly $10^{-5}\,\rm{GeV^{-1}}$ over a range of masses from 0.4 to 50 GeV.
\vfill\eject

\section {Introduction}
One of the simplest BSM scenarios comes from augmenting the Standard Model with new singlet (pseudo)scalar particles. Such models have rich phenomenology despite their simplicity, and can influence the structure of the Electroweak phase transition \cite{Carena:2019une}, provide natural dark matter candidates \cite{Duffy:2009ig,Athron:2017kgt}, and can be naturally accommodated in well motivated UV models \cite{Svrcek:2006yi}. One class of new light scalars is the Axion-Like Particle (ALP) \cite{Ringwald:2014vqa,Beacham:2019nyx}. An ALP is defined as a relatively light pseudo-scalar that couples to two gauge bosons and possibly SM fermions. Via the PQ mechanism \cite{PhysRevLett.40.223} or other tunings, these particles are particularly well motivated as solutions to the Dark Matter and Strong CP problems, but can appear generically as pseudo Nambu-Goldstone bosons of spontaneously broken approximate symmetries or descend from phenomenogical string theory models. Regardless of their origin, ALPs are an extremely general extension of the SM and serve as a test case for investigating BSM physics. In addition, as the LHC and other experiments search for new physics in the form of heavy ($ M \gg M_{W}$) particles, ALPs serve as an orthogonal but complementary search direction for theorists and experimentalists, as they are generally low mass but very weakly coupled. Fortunately, we will show that the next generation of lepton colliders like the ILC will provide clean signatures to weakly coupled ALP physics at the $\mathcal{O}(10^{-5}\,\rm{GeV})$ level for a range of interesting masses near and below the weak scale.
\vskip 0.12in
Many dedicated search strategies have been developed to study their production and influence on cosmology and particle physics, depending on the exact nature of the ALP in question \cite{Kim:1986ax}. In more detail, these searches depend on which gauge bosons the ALPs couple to, e.g. $U(1)_{Y}$, $SU(2)_{L}$, or $SU(3)_{C}$, and whether the the ALP couples to the SM fermions. In an effective field theory approach each of these couplings should be allowed, but each can be taken to be independent (modulo RG effects \cite{Bauer:2020jbp,Chala:2020wvs}), allowing one to examine each portal one at a time. In this paper we take up one ALP model where our Axion-Like Particle, $a$, couples only to hypercharge through a dimension 5 interaction,
\begin{equation}
{\cal L} = {\cal L}_{\rm SM}+\frac{1}{2}\partial_\mu a\partial^\mu a -\frac{1}{2}m^2_{a}a^2 - \frac{g_{a BB}}{4}a B_{\mu\nu}\tilde{B}^{\mu\nu}~~{\rm (representative~theory).}
\label{eq:theory}
\end{equation}
Here, $m_{a}$ is the tree level mass of the ALP which couples to hypercharge via the dimensionful coupling $g_{a BB}$, and $\tilde{B}^{\mu\nu} = 1/2\varepsilon^{\mu\nu\alpha\beta}B_{\alpha\beta}$ is the dual hypercharge field strength tensor. Of course many other effective Lagrangians involving couplings to additional gauge bosons and fermions would be equally valid to write down, but the resulting phenomenology is either qualitatively similar or produces orthogonal observables which would not affect subsequent discussion. Note, that several other authors study ALPs below the weak scale which couple purely to $F\tilde{F}$ rather than to $B\tilde{B}$. In the former case, the ALP couples only to photons and cannot couple to the $Z$. We choose to study the latter, $aB\tilde{B}$ operator, because in a gauge invariant, UV completion of any model including an ALP there are likely to be $\mathcal{O}(1)$ connections between operators coupling the ALP to each of the electroweak gauge bosons. Thus we would like to put ALP-photon couplings on equal footing with ALP-$Z/W$ couplings. This, as we shall see below, opens up further experimental discovery channels.
\section{ALPs in Rare Z Decays}
\begin{figure}[htbp]
\begin{center}
\includegraphics[width=16cm]{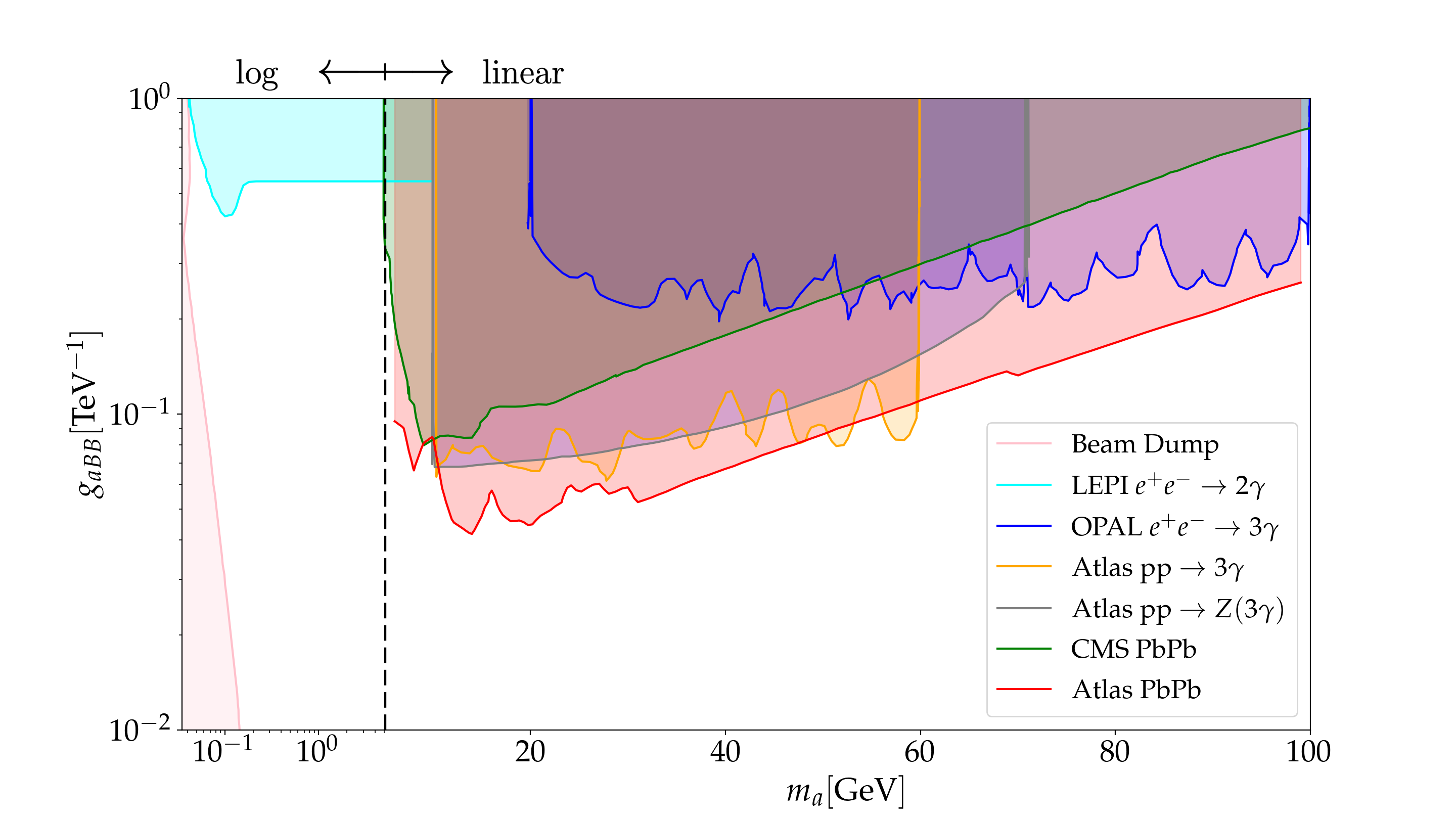}
\caption{Current constraints on ALP model with hypercharge coupling. Figure adapted from  \cite{Jaeckel:2015jla,Sirunyan:2018fhl,Aad:2020cje,Aad:2015bua,Knapen:2016moh}.}
\label{fig:current_constraints}
\end{center}
\end{figure}

\begin{figure}[t]
\begin{center}
\includegraphics[width=12cm]{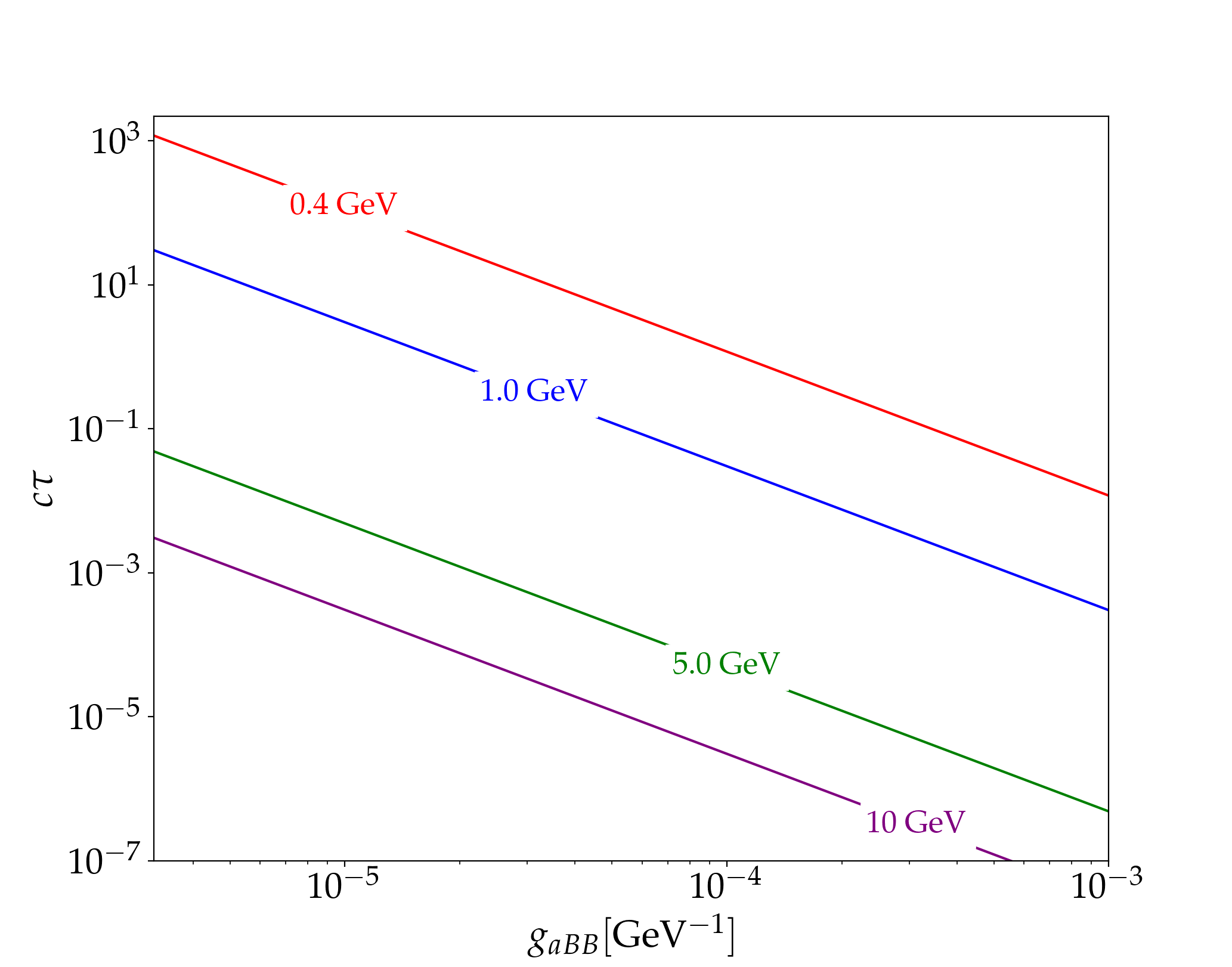}
\caption{Proper decay length as a function of $g_{aBB}$ for several different ALP masses.}
\label{fig:decay_length}
\end{center}
\end{figure}

Much of the parameter space of this particular model, shown in Fig.~\ref{fig:current_constraints}, is highly constrained by terrestrial experiments \cite{Graham:2015ouw}, as well as cosmology and astrophysics \cite{Powell:2016tfs,Millea:2015qra}. Light shining through wall (LSW) experiments and helioscopes constrain ALP masses up to several eV, and down to $g_{a BB} = 10^{-11}\,\rm{GeV}^{-1}$, while cosmology and astrophysical constraints cover larger masses in the eV to GeV range and couplings down to $10^{-12}\,\rm{GeV}^{-1}$ and lower.
Complimentary constraints can be obtained with colliders and beam dump experiments, which can probe masses in the MeV to TeV range. ALPs produced at beam dumps penetrate shielding and then decay to pairs of photons which are detected by a downstream detector. These rely on relatively smaller ALP couplings than colliders as the ALP has to travel a macroscopic distance to make it through the shielding and reach the downstream detector. Future experiments like FASER and DarkQuest will probe ALP-photon couplings around $10^{-3} - 10^{-6}\,\rm{GeV^{-1}}$ in  10 MeV - several 100 MeV range \cite{Feng:2018pew,Kowalczyk:2920agg}. Collider searches target ALP production in association with a photon, with the ALP either leaving the detector leading to MET, or decaying in the detector volume leading to displaced vertices or other identifying signatures. Belle II (not shown in Fig.~\ref{fig:current_constraints}) for example searches for ALPs in $e^{+} e^{-} \rightarrow a\gamma \rightarrow 3\gamma$, and has found constraints of $g_{a\gamma\gamma} < 10^{-3}\,\rm{GeV}^{-1}$ in the mass range of 0.2 GeV to 10 GeV \cite{BelleII:2020fag}.
\vskip 0.12in
Interesting constraints also come from light by light scattering at the LHC in PbPb collisions. Here, the $\gamma\gamma$ scattering cross sections are enhanced by a factor of $\text{Z}^{4}$ \cite{Knapen:2016moh}, where Z is the number of protons in the nucleus. The presence of an ALP which couples to photons would enhance the light by light scattering ($\gamma\gamma\rightarrow\gamma\gamma$) cross section. Measurements of this cross section at CMS and ATLAS place constraints on Axion-Like Particles coupling to photons from 5 to 100 GeV down to a few $\times10^{-4}$ GeV \cite{Sirunyan:2018fhl,Aad:2020cje,Knapen:2016moh}, competitive with LEP and other LHC searches over this mass range.
\vskip 0.12in
We would like to mention the analysis by Bauer et al. in \cite{Bauer:2018uxu} where the authors consider the ALP discovery prospects of the FCC-ee, HL-LHC, CLIC, and other future experiments in the context of a similar ALP model in the MeV to TeV mass range. Their analysis concludes that these experiments will probe ALP-photon couplings down to $10^{-6}\,\rm{TeV^{-1}}$ for certain ranges of masses. We caution that a detector level analysis with realistic cuts and a thorough background analysis will most likely weaken the claimed sensitivities.
\vskip 0.12in
After electroweak symmetry breaking the ALP develops a coupling to both the $Z$ boson and to the photon, opening up the decay of an on-shell $Z$ into an ALP and a photon with width
\begin{equation}
\Gamma_{Z\rightarrow a + \gamma} = g_{a BB}^2 s_{W}^{2}c_{W}^{2}\frac{(m_{Z}^{2} - m_{a}^{2})^3}{96\pi m_{Z}^3}.
\end{equation}
As long as the ALP has a mass less than $m_{Z}$, it will then decay to two photons with a branching ratio of nearly 1. It can of course decay to other standard model particles at loop level but these will be heavily suppressed. This leads to the decay chain $Z\rightarrow 3\gamma$ or "tri-photon" signature.
\vskip 0.12in
Whether the ALP will promptly decay or lead to a displaced vertex depends on the ALP mass and hypercharge coupling, with the decay length given by $l = c\tau = \gamma_{a}/\Gamma_{a \rightarrow \gamma\gamma}$, where $\gamma_{a} = E_{a}/m_{a}$ is the boost factor of the ALP. While dedicated searches for displaced vertices and or MET lead to interesting constraints \cite{Gershtein:2020mwi,Darme:2020sjf}, this note will focus on masses and couplings which lead only to prompt decays. This is shown in Fig.~\ref{fig:decay_length} for $m_{a} = 0.4, 1.0 , 5.0, 10$ GeV. For $m_{a} \geq 0.4$ GeV and for couplings which can be probed by the ILC, all decays are prompt (decays occur prior to ALP entering the ECAL at the ILC).
\vskip 0.12in
$Z\rightarrow 3\gamma$ is an interesting final state to search for as it is absent at tree level in the SM, but loop induced with a tiny branching ratio of $\approx 10^{-9}$ \cite{zphotonSM}. This rare decay has been searched for by LEP, the Tevatron, and the LHC in an effective field theory context searching for SM and anomalous $Z\gamma$ couplings \cite{Acciarri:1994gb, Aaltonen:2013mfa, Aad:2015bua}, with no excess over the SM background being found, establishing an upper limit of $B(Z\rightarrow 3\gamma) < 2.2\times10^{-6}$. Translating this into the $(m_{a}, g_{a BB})$ plane leads to the constraint $g_{a BB} < 10^{-4.5}\,\rm{GeV^{-1}}$, for masses large enough where the two photons from the ALP decay can be independently resolved. The exact value of the ALP mass, $m_{a}$, which leads to well separated decay photons depends on the details of the experimental analysis and detector and cannot straightforwardly be pinpointed, though it should lie at least above 10 GeV. Below $\mathcal{O}(10\,\rm{GeV})$ lies a subtle region of parameter space in collider searches which is the subject of this paper.
\vskip 0.12in
In dedicated ALP searches at experiments like the LHC and LEP, analyses are limited by their ability to reconstruct collimated pairs of photons \cite{Mimasu:2014nea}. Low mass ALPs produced from on-shell $Z$ decays, subsequently decay into pairs of photons with a $\Delta R$ separation which peaks at $4 m_{a}/m_{Z}$. Here $\Delta R= \sqrt{\Delta\phi^{2} + \Delta\eta^{2}}$ is the angular distance measure between particles. ALPs with masses between 1 and 10 GeV tend to decay to collimated photon pairs which will overlap in a detector and thus not be correctly reconstructed as two individual photons. See \cite{Sheff:2020jyw} for a discussion on reconstructing overlapping photons at the LHC. At the LHC and in the LEP experiment, photons must be separated from charged particles and other photons by a $\Delta R  > 0.2$, though this requirement depends on the details of each detector as well as the algorithms responsible for reconstructing photons. This should be taken as a rule of thumb. If a photon does not meet these requirements it is not reconstructed and rejected as a photon candidate. Above $m_{a} = 10$ GeV, the photons are well separated enough to be efficiently reconstructed. Low mass ALPs ($m_{a} < 1$ GeV) can be searched for in a two photon analysis as the two photons from the ALP decay tend to be collimated enough to register as one photon in a detector.
\vskip 0.12in
In \cite{Jaeckel:2015jla} this was used to recast the LEP search for $Z\rightarrow 3\gamma$ in the $e^{+}e^{-}\rightarrow \gamma\gamma(\gamma)$ measurement into limits on the ALP model. In different ALP mass regions, $m_{a} < m_{\pi^{0}}$, $m_{\pi^{0}} < m_{a} < 10$ GeV, and $m_{a} > 10$ GeV, either searches for two or three photons were used based on whether the photons from the ALP decay would be well separated or not. This was used to set leading bounds on ALP masses and couplings between 100 MeV and 90 GeV, with the weakest bounds being in the intermediate mass regime of 1 GeV to 10 GeV. Though jet substructure techniques can be used to disentangle collimate pairs of photons from single photons \cite{Ellis:2012zp}, an experiment that can resolve individual photons via a granular detector in this ALP mass regime is crucial to enhance discovery prospects and increase our sensitivity to new physics with the same decay topology.
\vskip 0.12in
The International Linear Collider (ILC) represents exactly this opportunity to search for this rare decay in the sought after intermediate mass regime. Recent advances in photon identification algorithms, combined with the highly granular ILC detectors can allow for photon identification with much more relaxed photon separation requirements, meaning photons can be much closer to other charged particles and other photons. In the next section we will investigate to what extent the ILC can improve on past searches for $Z\rightarrow a\gamma \rightarrow 3\gamma$.
\section{ILC and Photon Reconstruction}\label{section:ilcphotons}
\begin{figure}[htbp]
\begin{center}
\includegraphics[width=12cm]{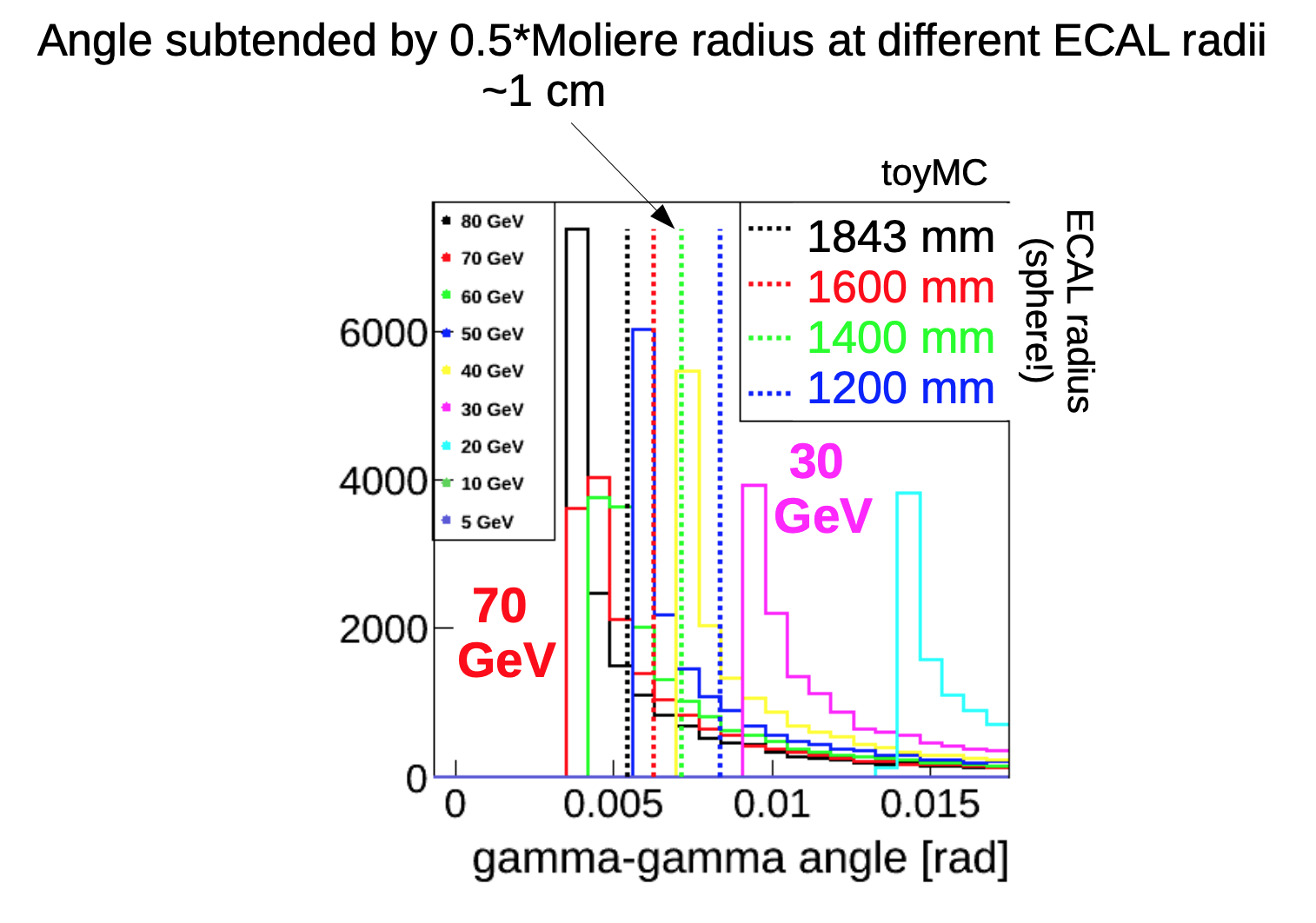}
\caption{Angle between photons from $\pi^{0}$ decays as a function of $\pi^{0}$ energy. Also depicted with dashed lines are the angle subtended by half a Moli\`ere radius at different ECAL radii. Figure from \cite{Jeans:2012jj}.}
\label{fig:photon_angles}
\end{center}
\end{figure}

The ILC is a next generation, high luminosity, linear $e^{+}e^{-}$ collider designed for high precision SM and BSM physics measurements \cite{Bambade:2019fyw, Erler:2000jg, Baer:2013cma}. The collider, though nominally designed to operate at 250 GeV center of mass energy, can be adjusted to run at a variety of center of mass energies, including operating at the $Z$ pole. Operating the ILC at $\sqrt{s} = m_{Z}$ is dubbed the Giga-Z mode of the ILC, as the ILC will produce on the order of $10^{9}$ or greater $Z$ bosons, orders of magnitude more than the LEP physics program. This will result in drastic improvements in precision $Z$ measurements, and measurements of $\text{sin}^{2}\theta_{\text{eff}}$. Production of this many $Z$ bosons will also allow for the search for rare decay modes, one of which is $Z\rightarrow 3\gamma$.
\vskip 0.12in
We first begin with a discussion of photon reconstruction at the ILC. Photon identification can be done via the GARLIC photon reconstruction algorithm \cite{Jeans:2012jj} which was developed by the International Large Detector (ILD) group. GARLIC (GAmma Reconstruction at a LInear Collider experiment) is designed to achieve highly efficient identification of photons within hadronic showers, which mostly come from high energy neutral pion decays. Because the photons from these pions will be highly collimated, this same technique can be used to identify collimated photons from low mass ALP decays. To begin, we examine in Fig.~\ref{fig:photon_angles} the angle between photons from $\pi^{0}$ decays as a function of the $\pi^{0}$ energy and the ECAL radius at the ILD (nominal ECAL radius is 1843mm).
The photon reconstruction performance is a function of the Moli\`ere radius, which is the transverse radius at which a single photon deposits 90\% of its energy. The smaller the Moli\`ere radius, the more separated each single photon will be. At $E_{\pi} = 20$ GeV, the Moli\`ere radius is roughly half the distance between the pair of photons. At a pion energy of 20 GeV, GARLIC reconstructs 2 photons correctly about 85\% of the time.
\vskip 0.12in
We would like to adopt this performance to reconstruct photons from low-mass ALP decays. To do so we first need to understand what is the minimum $\Delta R$ between photon pairs that we can expect to reconstruct. If we take the results from the 20 GeV pion seriously, the photons from this decay have a peak separation of $\Delta R = 4\times m_{\pi}/E_{\pi} = .027$. We still need to impose separation criteria on the photons. We can again use the pion reconstruction as a test case.
\begin{figure}[t]
\begin{center}
\includegraphics[width=10cm]{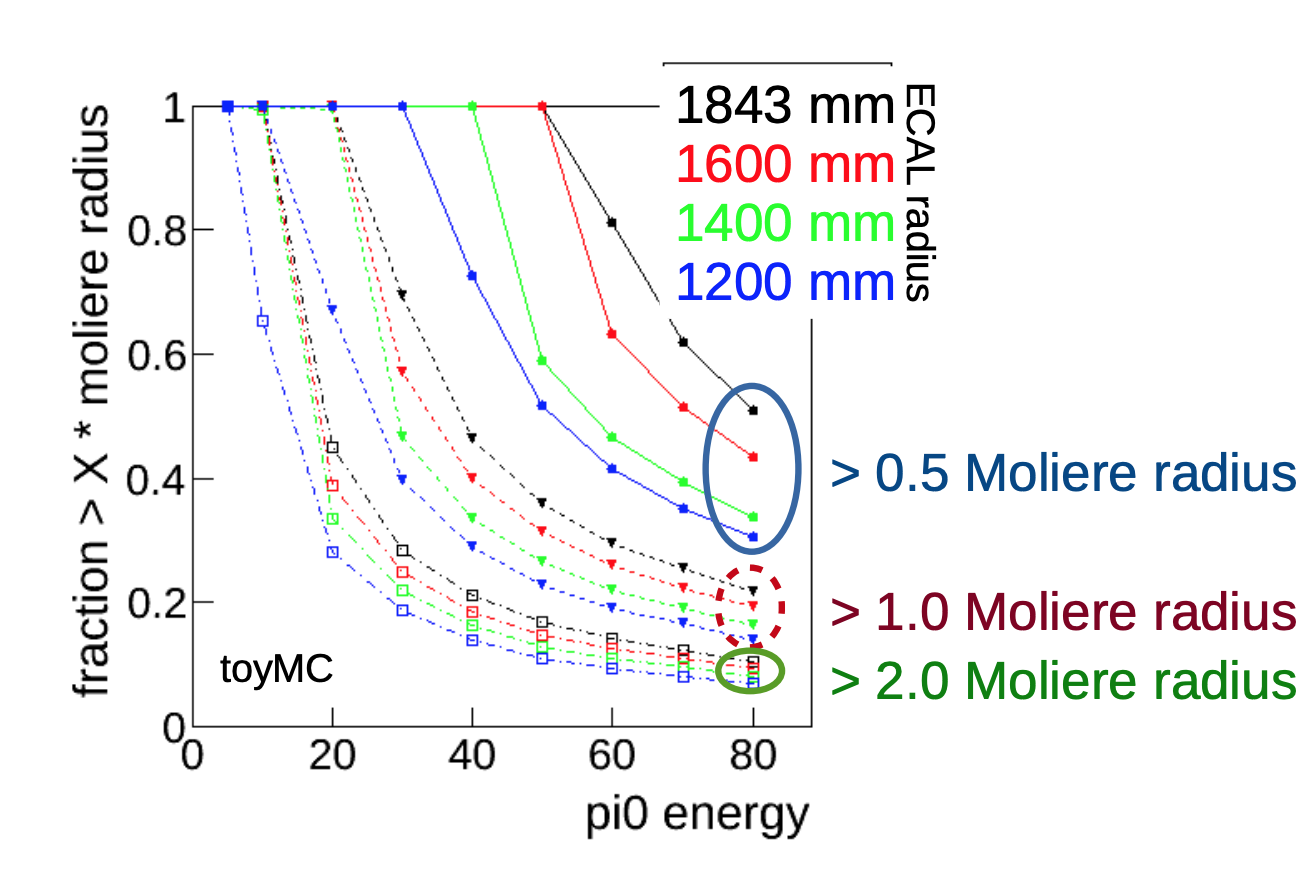}
\caption{Fraction of $\pi^{0}$ that have photons separated by greater than 2, 1, 0.5 Moli\`ere radius in the ECAL. Figure from \cite{Jeans:2012jj}.}
\label{fig:separation}
\end{center}
\end{figure}
Fig.~\ref{fig:separation} shows the fraction of $\pi^{0}$ that have photons separated by greater than 2, 1, 0.5 Moli\`ere radii in the ECAL. At the nominal ECAL radius, almost 100\% of pions with E = 15 GeV have photons separated by 2 Moli\`ere radii. This means that if we form cones around each photon of $\Delta R = 4\times m_{\pi}/(15\,\rm{GeV}) = .035$ \cite{Aaboud:2018djx}, then roughly 10\% of the energy inside each cone will be from the other photon. Thus we can use this as our photon separation criteria.
\vskip 0.12in
We choose to make photons with an isolation cone of $\Delta R = 0.035$, and a $PT_{iso} = 0.1$, meaning that no more than 10\% of the PT contained in our cone can come from another photon. The effect of this cut will have on our efficiency to identity photons can be seen from Fig.~\ref{fig:deltaRzdecay} below.
\begin{figure}[htbp]
\begin{center}
\includegraphics[width=14cm]{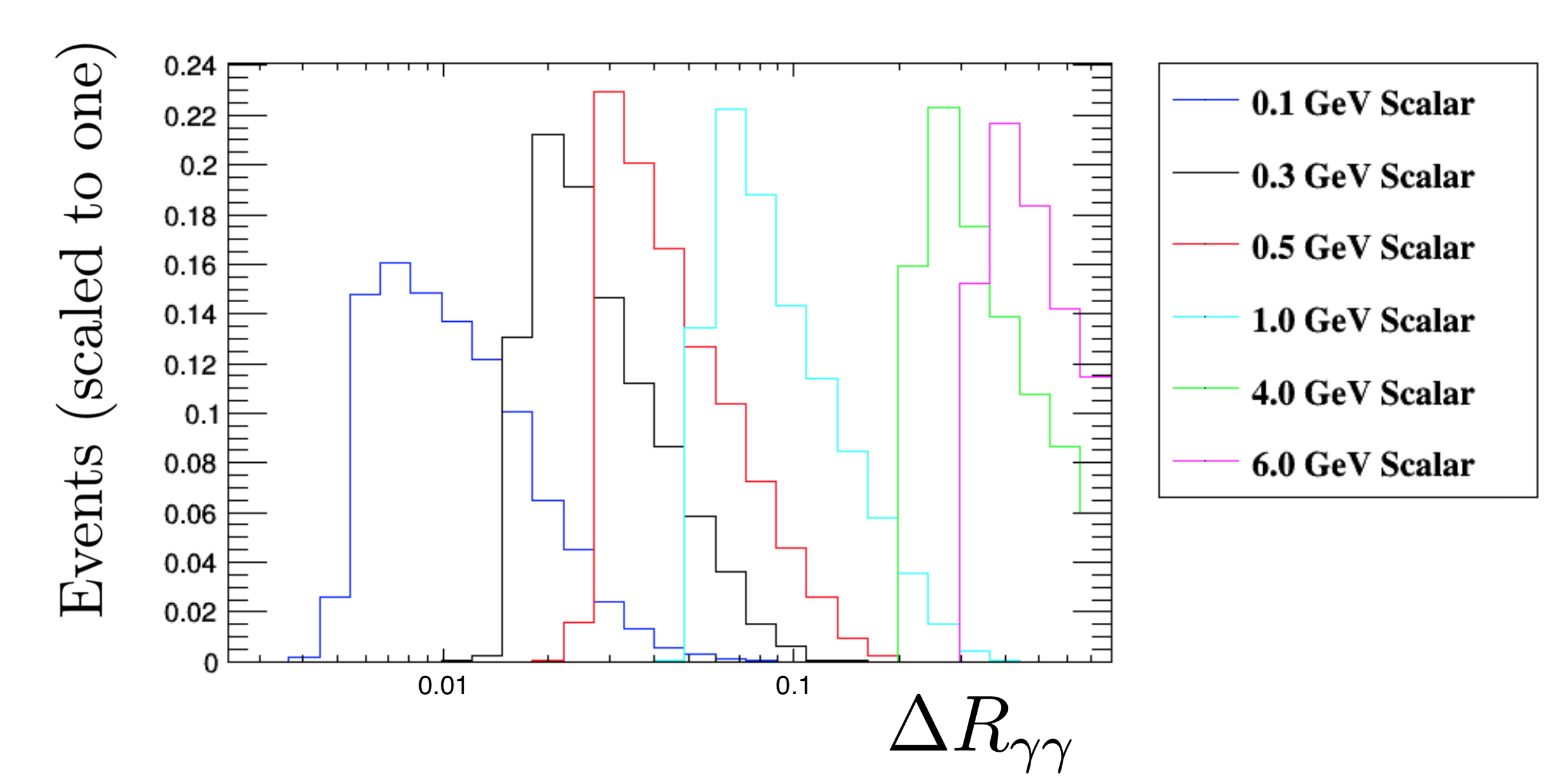}
\caption{$\Delta R$ between photons from the ALP decay for a range of ALP masses between 0.1 and 6.0 GeV. The peak of the distribution shifts towards higher $\Delta R$ for larger masses, following the form $\Delta R_{\rm{peak}} = 4m_{a}/m_{Z}$}.
\label{fig:deltaRzdecay}
\end{center}
\end{figure}
For masses below 0.5 GeV, the $\Delta R$ between photons can be significantly smaller than 0.035, making photon pair reconstruction challenging. At $m_{a}$ = 0.5 GeV, the peak of the distribution is near $\Delta R = 0.04$ which allows for sufficient separation for both photons in the pair. We simulated 50,000 $e^{+} e^{-} \rightarrow a \gamma \rightarrow \gamma\gamma\gamma$ events at $\sqrt{s} = m_{Z}$, for $m_{a} = {0.5, 1, 2, 3, 6}$ GeV. To test our separation criteria we can compute the average number of photons reconstructed. With perfect reconstruction we would expect this to be 3. Failing separation cuts, as well as failing a cut on the minimum photon energy (2 GeV), reduces this number. For $m_{a} = 0.5$ GeV we reconstruct on average 1.8 photons per event, significantly below 3. This is because quite often the pair of photons from the scalar decay fail separation criteria and thus are rejected. At $m_{a} = 1$ GeV and above the average number of photons reconstructed is 2.7 due to the much larger $\Delta R$ between photon pairs.
\section{Signal vs. Background}
\begin{figure}[t]
\begin{center}
\includegraphics[width=12cm]{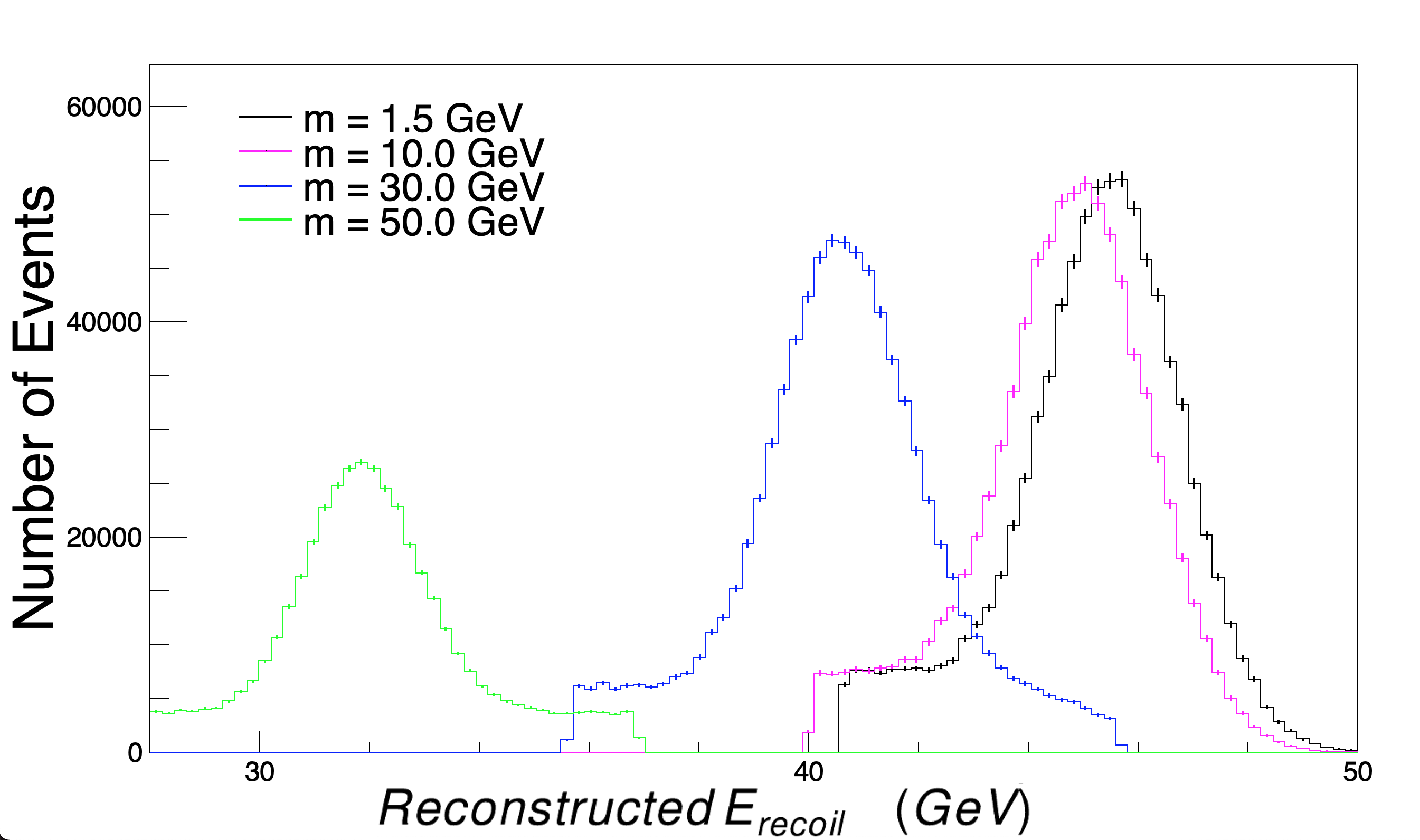}
\caption{Reconstruction recoil photon energy distribution from $Z\rightarrow a\gamma$ signal events. For each value of $m_{a}$ photons are required to have an energy within 5 GeV of $E^{\gamma}_{\rm recoil}(m_{a})$. Each reconstructed distribution is peaked near the true value of $E^{\gamma}_{\rm recoil}(m_{a})$.}
\label{fig:recoil_energy}
\end{center}
\end{figure}

\begin{figure}[t]
\begin{center}
\includegraphics[width=14cm]{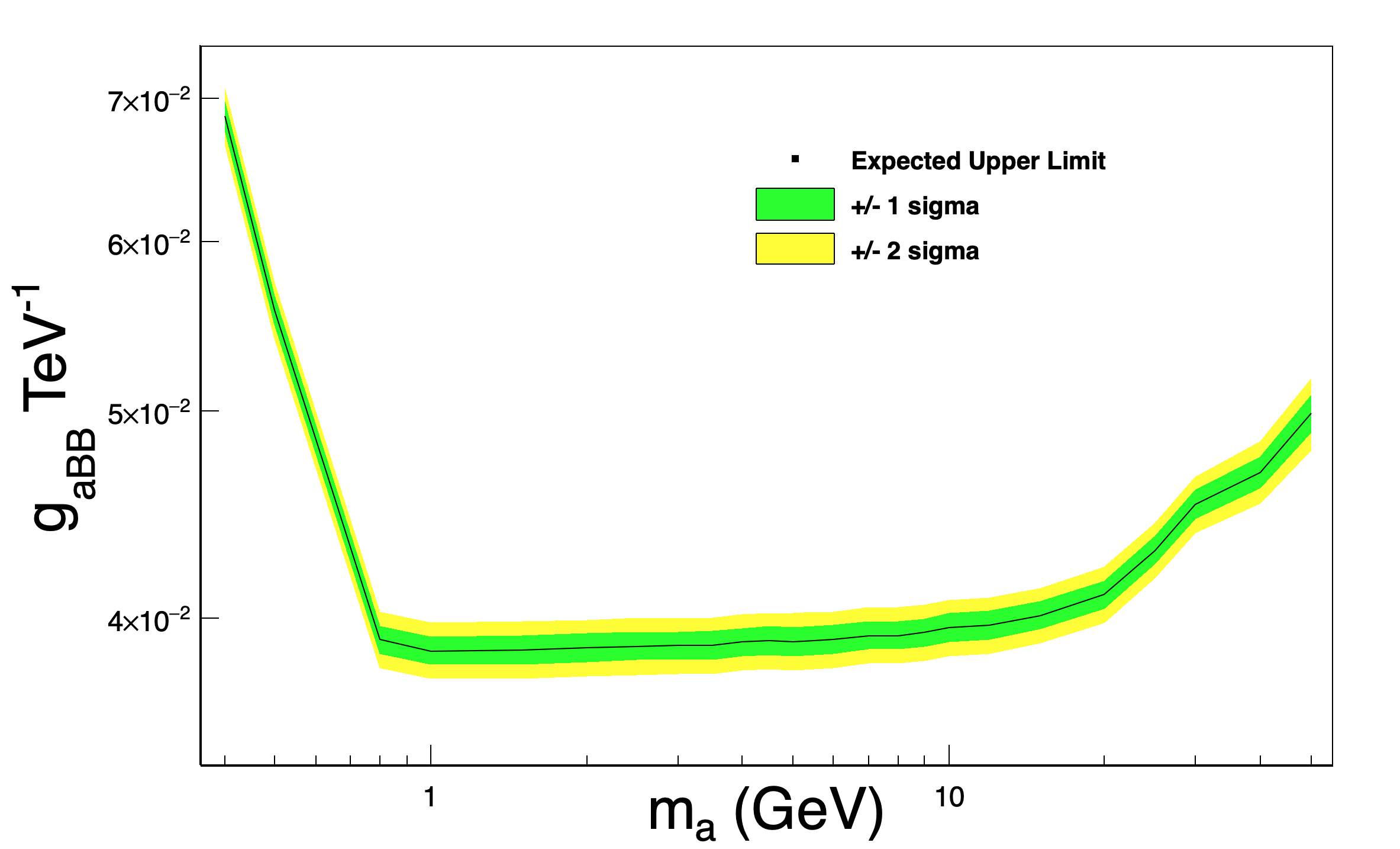}
\caption{Upper limit on $g_{aBB}$ over the range of masses 0.4 to 50 GeV. $1\sigma$ and $2\sigma$ error bars are shown in yellow and green.}
\label{fig:1d_exclusion}
\end{center}
\end{figure}

Searches for $Z\rightarrow 3\gamma$ have been made difficult because of the relatively large SM background $e^{+} e^{-} \rightarrow 3\gamma$, which has a cross section at $\sqrt{s} = m_{Z}$ of approximately 4.1 pb (computed at leading order using MadGraph aMC@NLO \cite{madgraph}).
\vskip 0.12in
The tree level $e^{+} e^{-} \rightarrow Z \rightarrow 3\gamma$ cross section in the ALP extension we are considering depends on the mass of the scalar, $m_{a}$,  and the coupling of the ALP to the hypercharge gauge bosons. We want to investigate the sensitivity to light scalars in this channel. We simulated 100,000 signal events over a range of masses from 0.4 GeV to 50 GeV using our signal model with MadGraph v2.6.7 aMC@NLO \cite{madgraph}, showered with Pythia 8.2 \cite{Sjostrand:2014zea}. We use the generic ILC Delphes card \cite{ILCDelphes} provided by the ILCSoft developers which simulates the response of a generic ILC detector. We modify the detector cards in accordance with section~\ref{section:ilcphotons}. To isolate our signal over the Standard Model background $e^{+}e^{-}$ background we need only make a small number of simple cuts. The first is of course that we have three non overlapping, efficiently reconstructable photons. The effect of this cut and its limitations are described in section~\ref{section:ilcphotons}. The second cut utilizes kinematic information from the two body, on-shell $Z$ decay. Being a two body decay, the energy of the recoiling photon is a fixed function of $m_{a}$, $E^{\gamma}_{\rm recoil}(m_{a}) = (M_{Z}^{2} - m_{a}^2)/2M_{Z}$. Thus, when searching for an ALP of mass $m_{a}$, one can require that one photon out of the three have an energy near $E^{\gamma}_{\rm recoil}(m_{a})$. We choose to search for photons within 5 GeV of this energy. These cuts are summarized below.
\begin{enumerate}
\item 3 non-overlapping ($\Delta R > .035$) photons with $E_{\gamma} > 2$ GeV
\item $|E_{\gamma} - E^{\gamma}_{\rm recoil}(m_{a})| < 5$ GeV
\end{enumerate}
For small $m_{a}$, $E^{\gamma}_{\rm recoil}$ is approximately $m_{Z}/2$. As can be seen in Fig.~\ref{fig:recoil_energy}, the recoil photon energy cut is especially import for $m_{a}$ above 30 GeV where the recoil energy starts to decrease significantly from $m_{Z}/2$.
Below in Fig.~\ref{fig:yield} we plot the signal and background yield as a function of $m_{a}$ with $g_{aBB} = 1\,\rm{TeV^{-1}}$. With just the above cuts, the ILC Giga-Z would observe $\mathcal{O}(10^{6}\,\rm{events})$ over the entire range of masses between 0.4 GeV and 50 GeV with only 100 $fb^{-1}$ of integrated luminosity.
\begin{figure}[htbp]
\begin{center}
\includegraphics[width=14cm]{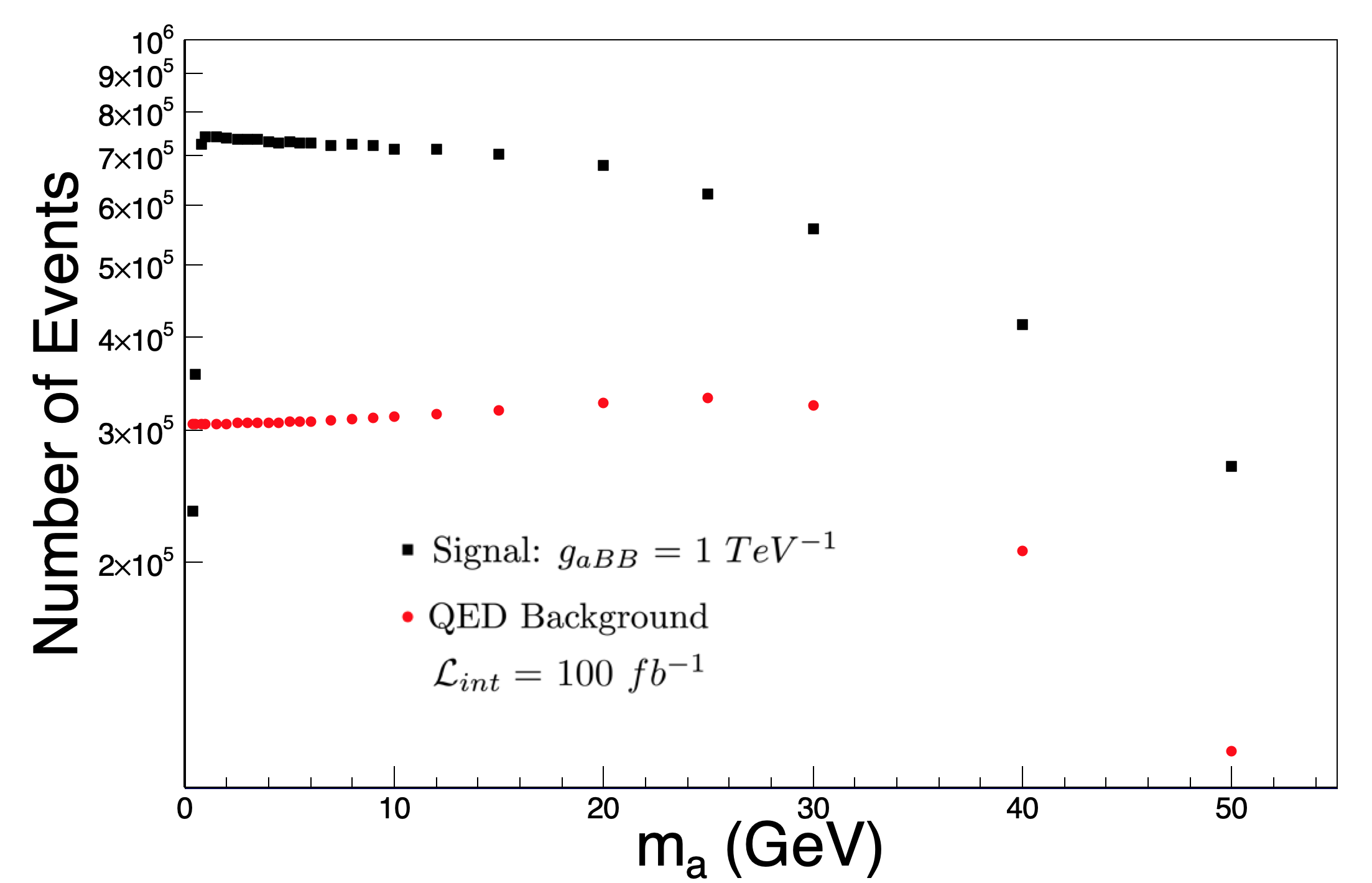}
\caption{Signal and background yields for a range of ALP masses for fixed $g_{aBB} = 1\,\rm{TeV^{-1}}$. Yields are normalized to an integrated luminosity of 100 $\rm fb^{-1}$. Almost $10^{6}$ signal events are expected for this value of $g_{aBB}$, well above the Standard Model background.}
\label{fig:yield}
\end{center}
\end{figure}
We can easily examine the sensitivity of the ILC to the above model as the coupling $g_{a BB}$ decreases. For a given coupling, $g$, producing a signal yield, $N$, we can find the signal yield, $N'$, with coupling $g'$ simply by using $N' = N(\frac{g'}{g})^{2}$. To produce an upper limit on $g_{aBB}$ at a 95\% confidence level, we can look for when the signal yield, as a function of $m_{a}$ and $g_{a BB}$, exceeds two times the uncertainty on the background. Essentially, if the background yield in the signal region for $m_{a} = m$ GeV is B events, then we can exclude all couplings for an ALP with mass $m$ GeV that produce more than $2\times\sqrt{B}$ signal events. Doing this for a range of masses gives us the exclusionary power of the ILC Giga-Z program at 95\% confidence.
\vskip 0.12in
After obtaining the signal and background yields in our mass range of interest, we compute the upper limit on $g_{aBB}$, shown in Fig.~\ref{fig:1d_exclusion}, according to the procedure described above. We include $1\sigma$ and $2\sigma$ error bars which take into account the statistical uncertainties on the signal yield. Obviously in a real analysis systematic experimental and theoretical uncertainties would have to be taken into account, but we do not expect this to modify our bounds a significant amount. We find that at masses between 0.4 and 50 GeV, the ILC can place stringent upper limits on $g_{aBB}$ which improve on LEP's by over an order of magnitude. To compare against current bounds we plot in Fig.~\ref{fig:money} the ILC Giga-Z exclusion region along with current constraints, both as a function of $g_{aBB}\,\rm{(TeV^{-1})}$ and as a function of $g_{aBB}^{-1}\,\rm{(TeV)}$, which is indicative of the scale of new physics which produces the operator $\mathcal{O} = aB\tilde{B}$.
\begin{figure}[t]
\begin{center}
\includegraphics[width=15cm]{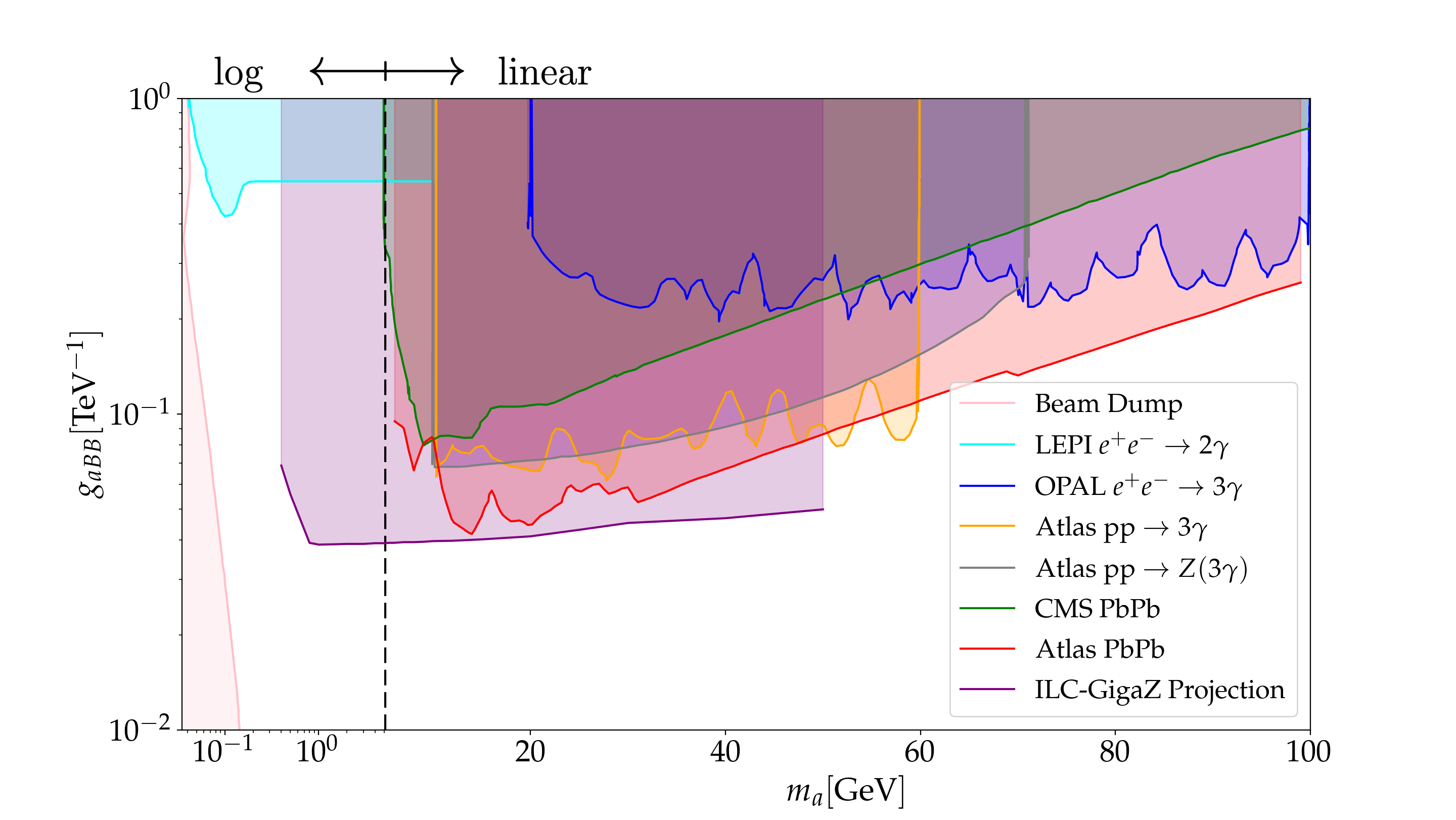}
\caption{ILC Giga-Z exclusion region against past experiments exclusion regions. We plot the limit here as a function of $g_{aBB}\,\rm{(TeV^{-1})}$ The ILC will significantly improve limits from over the whole range of masses from 0.4 to 50 GeV.}
\label{fig:money}
\end{center}
\end{figure}
\begin{figure}[t]
\begin{center}
\includegraphics[width=15cm]{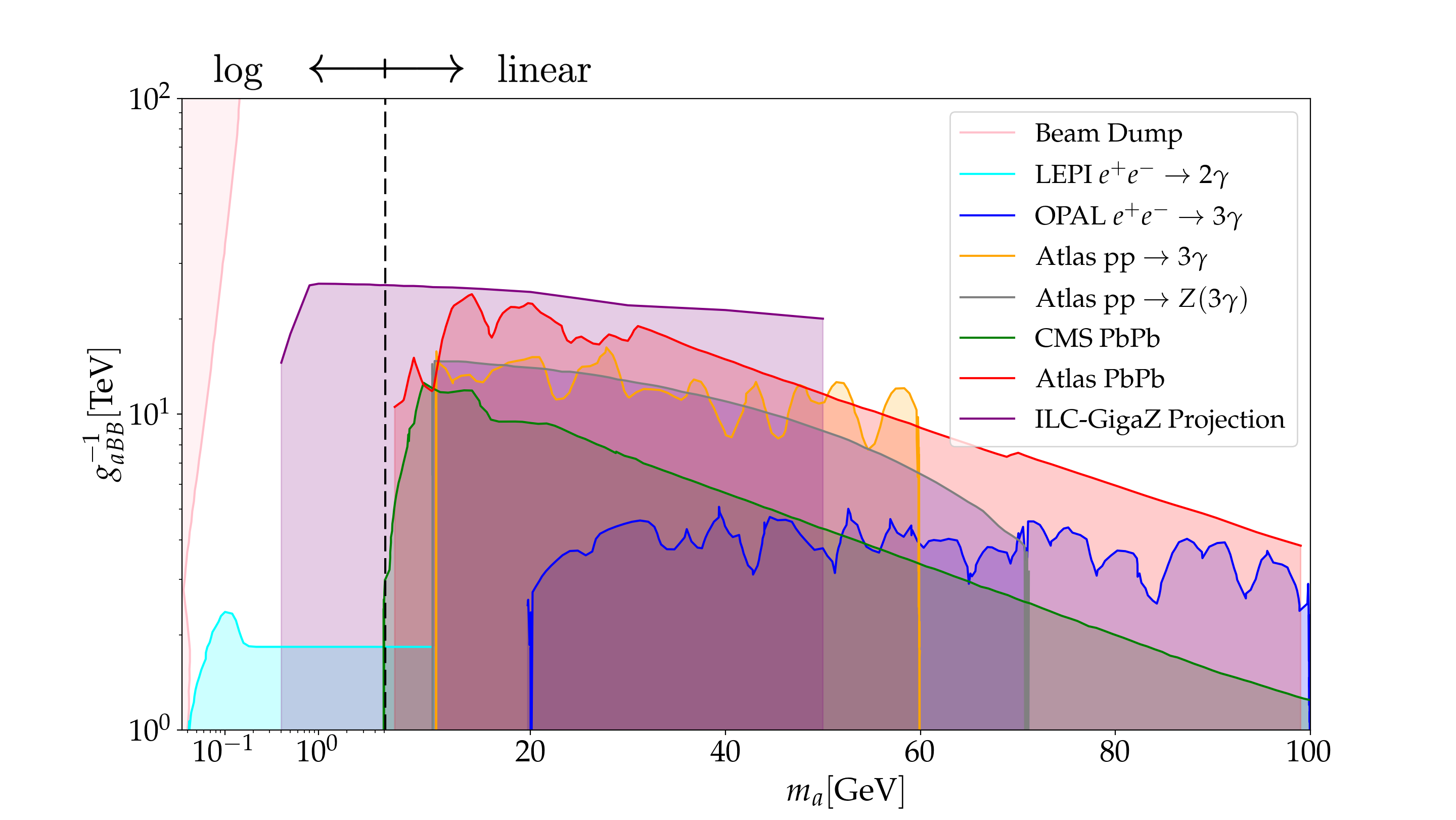}
\caption{ILC Giga-Z exclusion region against past experiments exclusion regions. We plot the limit here as a function of $g^{-1}_{aBB}\,\rm{(TeV)}$ The ILC will significantly improve limits over the whole range of masses from 0.4 to 50 GeV.}
\label{fig:money}
\end{center}
\end{figure}
We would like to stress that this simple analysis, based only on the granularity of the future ILC detectors and the kinematics expected from the $Z\rightarrow a\gamma$ decay, is able to probe much more deeply new regions of parameter space in this model. A more sophisticated experimental analysis would involve a bump-hunt search on the invariant mass of the two photon system from the ALP decay. Additional information which could be used to make these bounds even stronger could be obtained from the angular separation between the photons from the ALP decay. For $m_{a} < 20$ GeV, the distribution of $\Delta R$ of these two photons is peaked at low values (see Fig.~\ref{fig:deltaRzdecay}), while the background distribution falls quickly in this region. Utilizing the excellent angular separation abilities of the ILC detectors, one could require that this two photon system be below a particular angular separation, further reducing the background. Finally, for very small masses ($< 0.5\,\rm{GeV}$), the two photons from the ALP decay would be collimated enough as to overlap in the detector and possibly be registered as a single photon. Thus, a straightforward two photon analysis, or one which uses shower shape variables to discriminate overlapping photons from single photons could strengthen bounds in the low mass region.
\section{Conclusion}
Axion-Like Particles (ALPs) represent an exciting and generic, calculable SM extension, with a parameter space that is being investigated through many different fronts. We have shown that even small ($\mathcal{O}(10^{-2}\,\rm{TeV^{-1}})$) ALP couplings to hypercharge generate a significant number of signal events at an experiment like the ILC over a range of masses from 0.4 GeV to 50 GeV. The exceptional photon reconstruction abilities of the future ILC detector(s) will allow for efficient identification of 3 photon events even at very small photon separations. This represents an exciting opportunity for the ILC to probe new light, weakly coupled physics which interacts with the $Z$ and to make a discovery.
\section{Acknowledgements}
We thank Advanced Research Computing at the University of Michigan, Ann Arbor for their computational resources. This work was supported by the DOE under grant DE-SC0007859. N. Steinberg is supported by a fellowship from the Leinweber Center for Theoretical Physics.

\bibliographystyle{utphys}
\bibliography{ZDecay_Notes}

\end{document}